\documentclass[12pt]{article}
\pdfoutput=1
\usepackage{graphicx}
\usepackage{wrapfig}
\usepackage{color}
\usepackage{setspace}

\usepackage[colorlinks]{hyperref}
\hypersetup{ colorlinks,
  linkcolor=blue,
  filecolor=blue,
  urlcolor=blue,
  citecolor=blue}
\RequirePackage{xspace}
\def\lhcb {LHC{\em b\/}\xspace}
\newcommand\pubnumber{}
\newcommand\pubdate{\today}
\def\napoli{Department of Physics\\
Imperial College London, London SW7 2AZ, United Kingdom}

\def\Title#1{\begin{center} {\Large #1 } \end{center}}
\def\Author#1{\begin{center}{ \sc #1} \end{center}}
\def\Address#1{\begin{center}{ \it #1} \end{center}}

\newcommand\pubblock{\rightline{\begin{tabular}{l} \pubnumber\\
         \pubdate  \end{tabular}}}
\newenvironment{Abstract}{\begin{quotation}  }{\end{quotation}}
\newenvironment{Presented}{\begin{quotation} \begin{center} 
             PRESENTED AT\end{center}\bigskip 
      \begin{center}\begin{large}}{\end{large}\end{center} \end{quotation}}

\def\beq{\begin{equation}}
\def\eeq#1{\label{#1}\end{equation}}
\def\eeqn{\end{equation}}

\def\beqa{\begin{eqnarray}}
\def\eeqa#1{\label{#1}\end{eqnarray}}
\def\eeqan{\end{eqnarray}}

\let\bar=\overbar

\def\Dslash{\not{\hbox{\kern-4pt $D$}}}
\def\dslash{\not{\hbox{\kern-2pt $\del$}}}

\def\msb{{\bar{\ssstyle M \kern -1pt S}}}

\begin{document}
\begin{titlepage}
\pubblock

\vfill
\Title{Time-integrated measurements and prospects for the CKM angle $\gamma$ at \lhcb}
\vfill
\Author{Mike Williams\footnote{On behalf of the \lhcb collaboration}}
\Address{\napoli}
\vfill
\begin{Abstract}

The status and prospects of time-integrated measurements of the CKM angle $\gamma$ at \lhcb, the LHC's dedicated flavor physics experiment, are reviewed.  Yields obtained from early data taking are presented and extrapolations are made to estimate what can be expected to be obtained from the 2011 data.  The conclusions drawn from these extrapolations are that \lhcb will produce the world's best measurement of $\gamma$ by the end of 2011 and that the long-term outlook is excellent.

\end{Abstract}
\vfill
\begin{Presented}
6th International Workshop on the CKM Unitarity Triangle\\
University of Warwick, UK, September 6-10, 2010
\end{Presented}
\vfill
\end{titlepage}
\def\thefootnote{\fnsymbol{footnote}}
\setcounter{footnote}{0}

\section{Introduction}

In the Standard Model of Particle Physics, transitions between the three generations of quarks are described by the Cabibbo-Kobayashi-Maskawa (CKM) matrix.  The parameter in this matrix responsible for $CP$ violation is denoted by $\gamma$.  Time-integrated measurements of $\gamma$ from tree-level decays have been made using various $B \to D^{(*)}K^{(*)}$ decays at the $B$ factories; the combined sensitivity is $\sim20^{\circ}$~\cite{ref:ckmfit} ({\em n.b.}, the motivations for and current status of $\gamma$ measurements are discussed in detail elsewhere in these proceedings~\cite{ref:zupan,ref:anton}).  By the end of 2011, the \lhcb experiment at the LHC will already have recorded several orders of magnitude more $B$-meson decays than the $B$ factories.  This note will explore the prospects for making time-integrated measurements of $\gamma$ at \lhcb using this data. The data presented in these proceedings is from the entire 2010 \lhcb data set, not only what was available at the time of the conference.  It is also worth noting here that time-dependent $\gamma$ measurements are also being pursued at \lhcb (see, {\em e.g.}, Ref.~\cite{ref:tdgam1}).

\section{Measuring $\gamma$ from $B \to DX$ Decays at \lhcb}

\begin{figure}
 \begin{center}
  \includegraphics[width=0.4\textwidth]{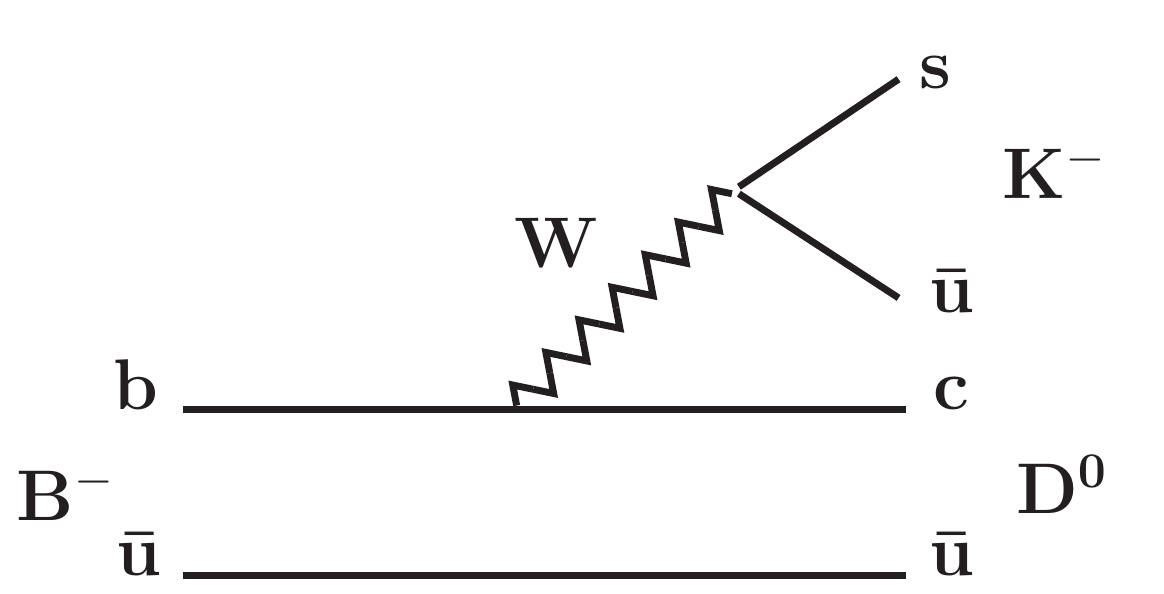}
  \hspace{0.05\textwidth}
  \includegraphics[width=0.4\textwidth]{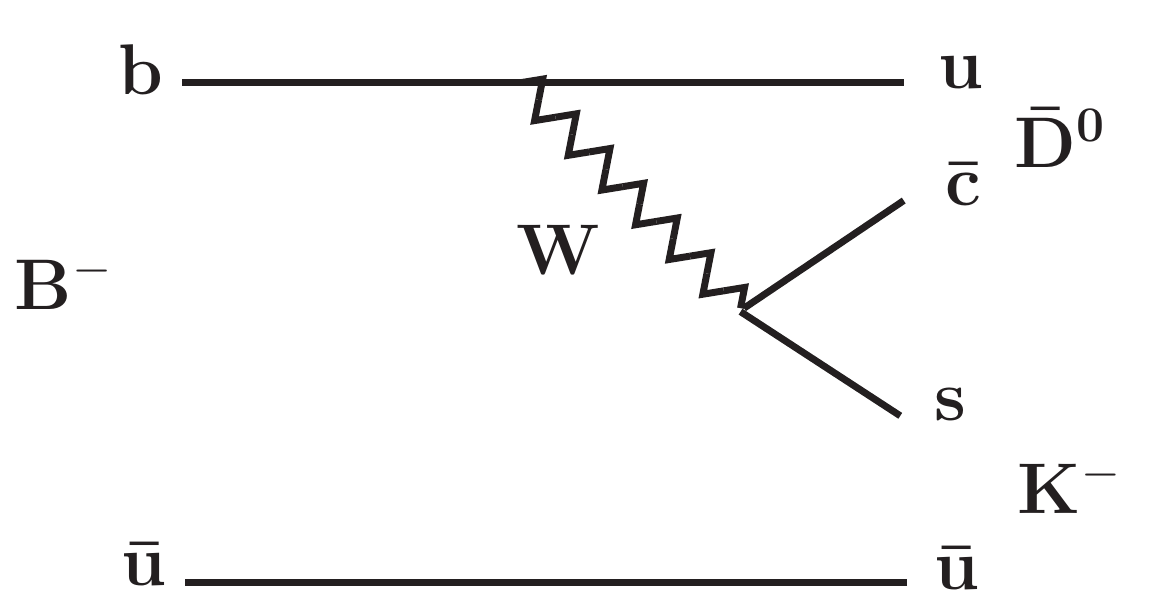}
  \end{center}
  \caption[]{\label{fig:diagrams1} Feynman diagrams for (left) $B^- \to D^0K^-$  and (right) $B^- \to \bar{D}^0 K^-$.}
\end{figure}

\begin{figure}
 \begin{center}
  \includegraphics[width=0.4\textwidth]{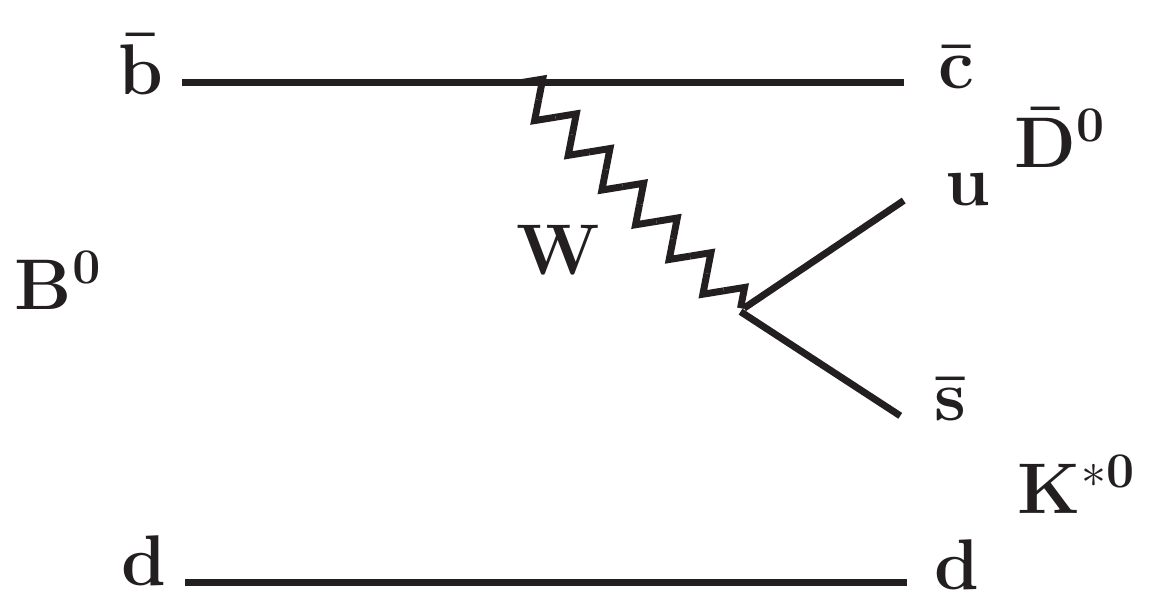}
  \hspace{0.05\textwidth}
  \includegraphics[width=0.4\textwidth]{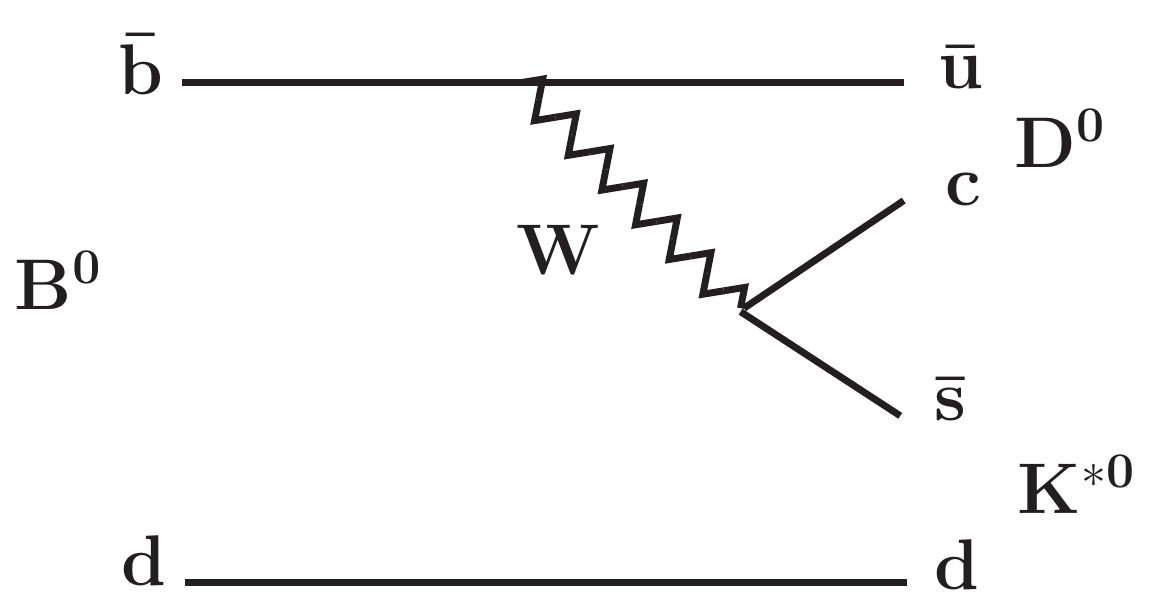}
  \end{center}
  \caption[]{\label{fig:diagrams2}
    Feynman diagrams for (left) $B^0 \to \bar{D}^0K^{*0}$  and (right) $B^0 \to D^0 K^{*0}$.
  }
\end{figure}

Gronau, London and Wyler (GLW) first proposed a method to measure $\gamma$ with negligible theoretical uncertainty using $B \to DK$ decays, where the neutral $D$ mesons are reconstructed in $CP$ eigenstates~\cite{ref:glw}.  The method was extended by Atwood, Dunietz and Soni (ADS) to use $D$ meson decays to any final state that is accessible to both the $D^0$ and $\bar{D}^0$ mesons~\cite{ref:ads}.  The sensitivity to $\gamma$ arises due to the interference between processes involving $b \to c\bar{u}s$ and $b\to u\bar{c}s$ transitions (see Figs.~\ref{fig:diagrams1} and \ref{fig:diagrams2}).
These methods are applicable to decays of both charged and neutral $B$ mesons.  

\begin{figure}
\begin{center}
  $B^{\pm} \to D_{\rm fav}(K\pi) K^{\pm}$ \hspace{1.5in} $B^{\pm} \to D(KK) \pi^{\pm}$ \\
  \includegraphics[width=0.49\textwidth]{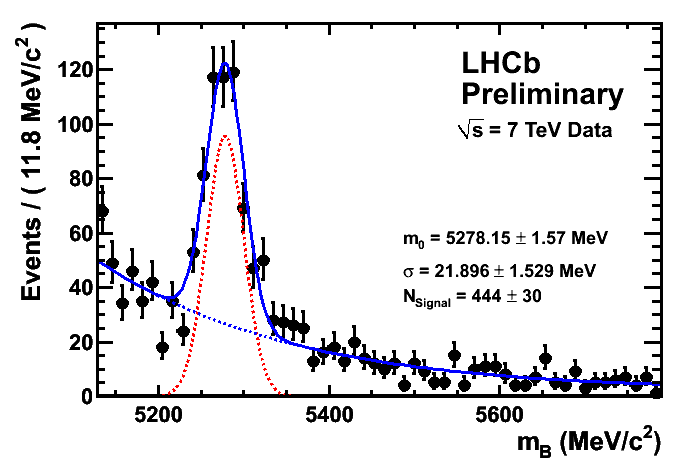}
  \includegraphics[width=0.49\textwidth]{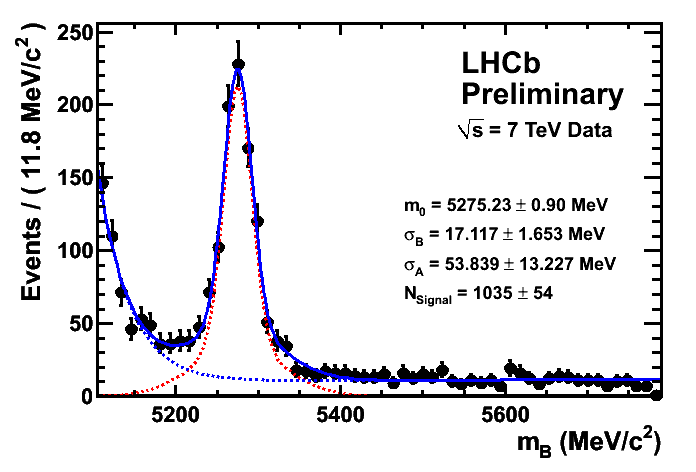}
\end{center}
\caption[]{\label{fig:masses1}
  Invariant mass distributions for (left) $B^{\pm} \to D_{\rm fav}(K\pi) K^{\pm}$ candidates and (right) $B^{\pm} \to D(KK) \pi^{\pm}$ 
candidates obtained from 34~pb$^{-1}$ of \lhcb data.
}
\end{figure}

The main decays of charged $B$ mesons that will be used to extract $\gamma$ using the combined GLW/ADS method at \lhcb are $B^{\pm} \to D_{\rm fav}(K\pi)K^{\pm}$, $B^{\pm} \to D_{\rm sup}(K\pi)K^{\pm}$ and $B^{\pm} \to D_{CP+}(KK,\pi\pi)K^{\pm}$.  Figure~\ref{fig:masses1} (left) shows the invariant mass distribution for $B^{\pm} \to D_{\rm fav}(K\pi)K^{\pm}$ candidates obtained from 34~pb$^{-1}$ of \lhcb data.  A total of $444\pm30$ signal events are observed.  One would then expect approximately 13000 $B^{\pm} \to D_{\rm fav}(K\pi)K^{\pm}$ events in 1~fb$^{-1}$, the expected amount of data to be collected by \lhcb by the end of 2011; however, for various reasons, the data taken in 2010 at \lhcb were collected using trigger settings that resulted in lower-than-nominal efficiencies.  If this is not the case in 2011, then the projections presented in these proceedings will underestimate the signal yields by as much as 30\%.  If the trigger efficiency is the same in 2011 as it was in 2010, then the resolution on $\gamma$ obtained at \lhcb using only $B^{\pm} \to D(hh)K^{\pm}$ decays will be comparable to the current combined sensitivity obtained by the $B$ factories from all $B \to D^{(*)}K^{(*)}$ decays.  \lhcb may also be able to improve its sensitivity by including $D \to K3\pi$ decays into the analysis.

\begin{figure}
\begin{center}
  $B^{\pm} \to D(K_S KK) \pi^{\pm}$ \hspace{1.5in} $B^{\pm} \to D(K_S\pi\pi) \pi^{\pm}$ \\
  \includegraphics[width=0.49\textwidth]{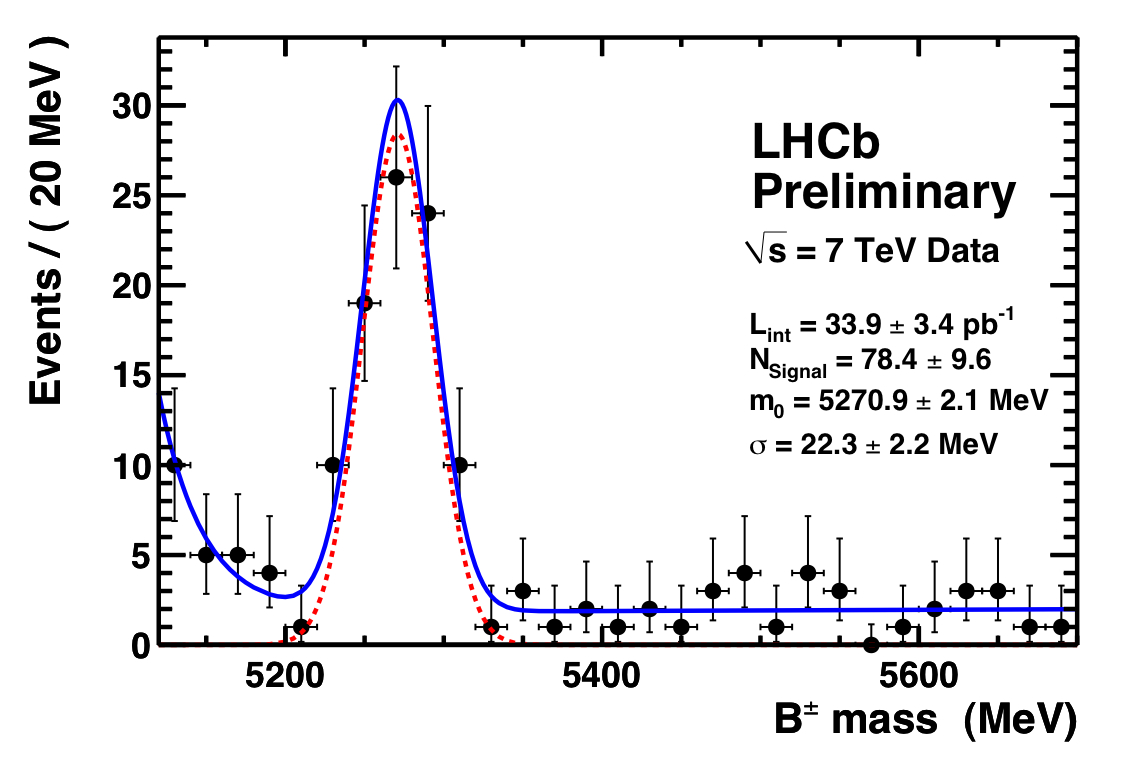}
  \includegraphics[width=0.49\textwidth]{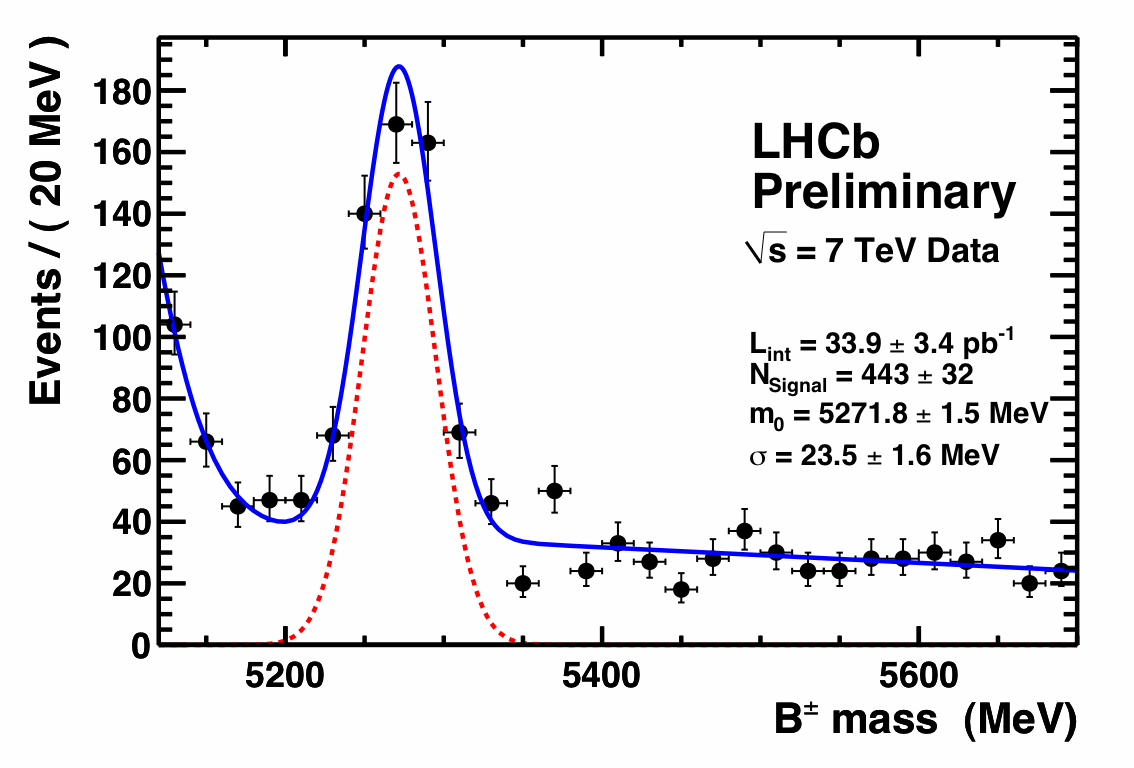}
\end{center}
\caption[]{\label{fig:masses2}
  Invariant mass distributions for (left) $B^{\pm} \to D(K_S KK) \pi^{\pm}$ candidates and (right) $B^{\pm} \to D(K_S\pi\pi) \pi^{\pm}$ candidates obtained from 34~pb$^{-1}$ of \lhcb data.
}
\end{figure}

The most precise $\gamma$ measurements made by the $B$ factories have been in the channels $B^{\pm} \to D(K_S KK)K^{\pm}$ and $B^{\pm} \to D(K_S \pi\pi)K^{\pm}$.  It is possible to extract $\gamma$ from these decays using both model-independent  and model-dependent methods~\cite{ref:ggsz}.  Figure~\ref{fig:masses2} shows the invariant mass distributions for the related decays $B^{\pm} \to D(K_S KK)\pi^{\pm}$ and $B^{\pm} \to D(K_S \pi\pi)\pi^{\pm}$ obtained from 34~pb$^{-1}$ of \lhcb data.  Accounting for the ratio of the $B^{\pm}\to DK^{\pm}$ and $B^{\pm}\to D\pi^{\pm}$ branching fractions, one would then expect a total of approximately 1000 $B^{\pm} \to D(KK+\pi\pi)K^{\pm}$ events in 1~fb$^{-1}$ (of course, the same caveats about the trigger efficiency apply to this projection as well).  This projects to an expected resolution on $\gamma$ at \lhcb from these channels alone of $\sim 16^{\circ}$.  


\begin{figure}
\begin{center}
  \includegraphics[width=0.49\textwidth]{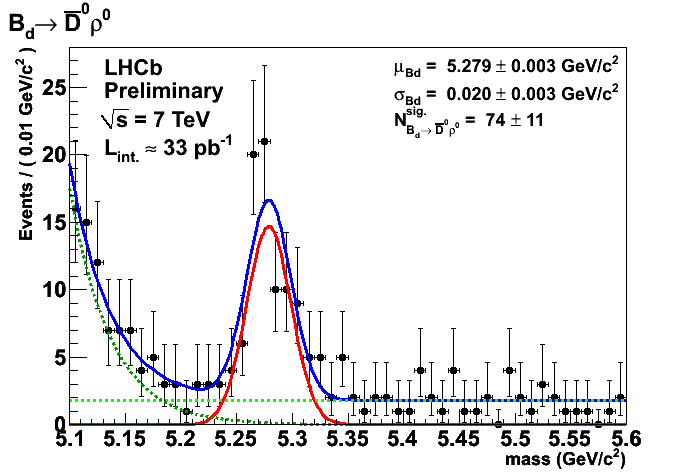}
\end{center}
\caption[]{\label{fig:masses3}
  Invariant mass distribution for $B^0 \to D_{\rm fav}(K\pi) \rho^0$ candidates obtained from 34~pb$^{-1}$ of \lhcb data.  
}
\end{figure}

Neutral $B$ mesons can also be used in the GLW/ADS methods.  For example, the decays $B^0 \to D(hh)K^{*0}$ can be used in the same way as the $B^{\pm} \to D(hh)K^{\pm}$ decays discussed above (with a minor modification to account for the non-zero width of the $K^{*0}$).  The sensitivity to $\gamma$ is enhanced in these decays due to the fact that both Feynman diagrams are color suppressed (see Figure~\ref{fig:diagrams2}); however, the statistics are lower which has the opposite effect on the resolution.  Figure~\ref{fig:masses3} shows the invariant mass distributions for the related decay $B^0 \to D_{\rm fav}(K\pi)\rho^0$ obtained from 34~pb$^{-1}$ of \lhcb data.  Accounting for the ratio of the $B^0 \to DK^{*0}$ and $B^0 \to D\rho^0$ branching fractions, one would then expect a total of approximately 300 $B^0 \to D_{\rm fav}(K\pi) K^{*0}$ events in 1~fb$^{-1}$ (with the same caveats as above).  This would not be enough statistics to make a $\gamma$ measurement competitive with the other methods discussed above; however, the resolution on $\gamma$ obtained from neutral $B$ decays can be greatly enhanced by considering the entire Dalitz plane using $B^0 \to D(hh)K\pi$ decays~\cite{ref:metim}.  A similar projection results in an expected yield of approximately 700 $B^0 \to D_{\rm fav}(K\pi)K\pi$ decays at \lhcb in 1~fb$^{-1}$.  This would provide a resolution on $\gamma$ of $\sim 20^{\circ}$.  Of course, the selections used to obtain these data sets are still evolving and future improvements are expected.

\section{Summary and Outlook}

\lhcb has already nearly reached its expected nominal offline selection efficiency and resolution.  The trigger efficiency was approximately 30\% lower than nominal in 2010; however, this is understood and may increase in 2011.  The long-term outlook for time-integrated measurements at \lhcb is excellent.  By the end of 2011, \lhcb will be able to produce the world's best measurement of $\gamma$.  Extrapolating further into the future carries many uncertainties, but what is certain is that \lhcb will significantly improve the world's knowledge of the CKM angle $\gamma$ in the next few years.

\begin{spacing}{0.5}

\end{spacing}
 
\end{document}